\documentclass{article} 
\usepackage{graphicx}        
\begin{document}
\centerline{\huge \bf A Biased Review of Sociophysics}
\medskip

\noindent
Dietrich Stauffer

\medskip

\noindent
Institute for Theoretical Physics, Cologne University,
D-50923 K\"oln, Euroland

\bigskip
{\bf Abstract.} Various aspects of recent sociophysics research are shortly 
reviewed: Schelling model as an example for lack of interdisciplinary
cooperation, opinion dynamics, combat, and citation statistics as an example
for strong interdisciplinarity. 

\section{Introduction}

The idea of applying physics methods to social phenomena goes back centuries
ago, e.g. with the first (unsuccessful) attempt to establish mortality tables,
involving astronomer Halley, the ``sociology'' of Auguste Comte who taught
analysis and mechanics around 1840, or the 1869 book by Quetelet ``Physique 
Sociale''. Majorana suggested to apply quantum physics in 1942 \cite{majorana}.
Some contemporary physicists \cite{weidlich,galam} have worked on the field 
since some decades. But it became a physics fashion about a dozen years ago, 
with opinion dynamics, applications of complex networks, etc. \cite{schweitzer}.
Presumably the best review is still the 
one from Italy during Berlusconi's rule \cite{fortunato}, which killed the
present author's chances to get a Nobel prize (in literature: science fiction)
for his four articles in \cite{granada} on languages (p. 49), opinions (p. 56). 
retirement demography (p. 69) and Bonabeau hierarchies (p. 75). The field is now
far too wide to be covered in a short review, and thus here only some biased 
selection is presented. Lecture notes of Fortunato \cite{forthist} start with
a nice introduction into the more ancient history of sociophysics, Galam wrote
a recent book \cite{galambook}, with pages 75-77 on: The Soviet-Style Rewriting 
of the History of Sociophysics.

We start with a discussion why it may be useful to apply physics research style 
to human beings, then we bring the Schelling model as an example where 
sociophysics was lacking for decades. The following three sections review 
opinion dynamics, combat, and citations, followed (for readers outside 
statistical physics) by a critique of mean field theory.  Econophysics is 
regarded here as outside of sociophysics, and also ignored because of recent 
reviews are languages \cite{ssw,acs,newbook}, Penna ageing models 
\cite{newbook,penna}, networks \cite{cohen} and traffic jams \cite{schad}.

\section{Does Sociophysics Make Any Sense?}

People are not atoms. We may be able to understand quite accurately the 
structure of the hydrogen atom, but who really understands the own marriage
partner. Nevertheless already Empedokles in Sicily more than two millennia 
ago found that people are like fluids: some people mix easily like water and 
wine (an ancient Greek crime against humanity), and some like water and oil
do not mix. And a few months ago the German historian Imanuel Geiss died, who
described the decay of empires with Newton's law of gravitational forces (but
disliked simulations to explain diplomatic actions during the few weeks before
World War I). 

It is the law of big numbers which allows the application of statistical
physics methods. If we throw one coin we cannot predict how it will fall, and
if we look at one person we cannot predict how this person will vote, when
it will die, etc. But if we throw thousand coins, we can predict that about 
500 will fall on one side and the rest on the opposite side (except when we
cheat, or the coin sticks in the mud of a sport arena). And when we ask 
thousand randomly selected people we may get a reasonable impression for an 
upcoming election. Half a millennium ago, insurance against loss of ships 
in the Mediterranean trade became possible, and life insurance relies on
mortality and the Gompertz law of 1825 that the adult probability to die within 
the next year (better: next month \cite{gavr} ) increases exponentially with 
age. Such insurance is possible 
because it relies on the large number of insured people: Some get money 
from the insurance and most don't. My insurance got years ago most of my savings
and now pays me a monthly pension until I die; the more the journal referees
make troubles to my articles, the sooner I die and the less loss my insurance
company will make with me. Only when all the banks and governments are coupled
together by their debts, the law of large numbers no longer is valid since
they all become one single unit \cite{aleks}. (``Maastricht'' rules on 
sovereign debts were broken by governments in Euroland since 1998, a decade 
before the Lehman crash.)   

Outside physics the method of agent-based computer simulations is fashionable
\cite{billari}. It has nothing to do with 007 James Bond, but refers to methods 
simulating single persons etc instead of averaging over all of them. Physicists
do that at least since 1953 with the Metropolis algorithm of statistical
physics, also in most of the simulations listed here.  This book \cite{billari} 
by non-physicists, written about simultaneously and independently from one by 
physicists \cite{newbook}, covers similar fields and similar methods but barely 
overlaps in the references. Recent work from cognitive science and related 
disciplines is listed in  \cite{cog}. 

Did sociophysics have practical applications?  When I got the work of Galam, 
Gefen and Shapir \cite{ggs}, I told
a younger colleague that I liked it. But after reading it he disliked it and 
remarked to me that the paper helps management to control its workers better
if a strike is possible. In the three decades since then I had read about 
many strikes but not about any being prevented by this paper. Two decades later
Galam \cite{chachacha}  was criticised by other physicists 
for having helped terrorists with his percolation application to terror 
support. I am not aware that such percolation theory was applied in practice.
However, a century ago physicist did not believe that nuclear energy can be
used.  Our own subsection ``Retirement Demography'' in ch. 6 of \cite{newbook} 
recommends immigration and higher retirement ages to balance the ageing of 
Europe; both aspects are highly controversial but I was not yet murdered.
Helbing's \cite{galambook} simulation that a column before a door improves
the speed of evacuation during a panic seems to me very practical.

\section{Schelling Model for Urban Ghettos}

The formation of urban ghettos is well known in the USA and elsewhere. Harlem
in Manhattan (New York) is the most famous ``black'' district since nearly 
one century, extending over dozens of blocks in north-south direction. Was
it formed by conscious discrimination e.g. from real-estate agencies, or was
it the self-organised result of the preferences of residents to have neighbours
of the same group? Of course, the Warsaw Ghetto, famous for its 1943 uprising,
was formed by Nazi Germany. Four decades ago, Schelling \cite{schelling} 
showed by a simple Monte Carlo simulation (by hand, not by computer) of two
groups A and B, that a slight preference of A people to have A neighbors, and
of B people to have B neighbors, suffices to form clusters of predominantly
A and predominantly B on a square lattice with some empty sites, out of an
initially random distribution. 

Statistical physicists of course would think of the standard Ising model on
a square lattice to understand such a question, with A people corresponding to
up spins and B people to down spins. Their ferromagnetic interaction gives
a preference of A for A neighbours, and the same for B. The temperature 
introduces randomness. Simulations with Kawasaki dynamics (conserved 
magnetisation) have been made since decades. However, it took three decades 
before physicists took up the Schelling  model; see  \cite{ortmanns,sumour9} for
an early and some  recent (physics) publications. 

It seems quite trivial that the equilibrium distribution is no longer random 
if people select their residence with A/B preferences; but does it lead to
``infinite'' ghettos, i.e. to domains which increase in size to infinity if the 
studied lattice tends to infinity? This phase separation is well studied in the 
two-dimensional Ising model: for temperatures $T$ above a critical temperature 
$T_c$, only finite clusters are formed. For temperatures below
this critical temperature, one large domain consisting mainly of group A, 
and another large domain consisting mainly of group B, are formed after 
a simulation time proportional to a power of the lattice size. 

Schelling could not see that his model does not give large ghettos, only small
clusters \cite{schelling} as for $T > T_c$ in Ising models,, but Jones 
\cite{jones} (from a sociology institute,
publishing in a sociology journal) corrected that by introducing more randomness
into the Schelling model; and Vinkovic (astrophysicist) and Kirman (economist)
did it two decades after Jones. Then large ghettos are formed as for Ising
models for $T < T_c$, Nevertheless, the Jones paper  today is cited 
much more rarely than Schelling, and mostly by physicists, not by his sociology
colleagues (see www.newisiknowledge.com for the Science Citation Index). Only
now the physics and sociology communities show some cross-citations for the 
Schelling model. 

Of course, instead of merely two groups A and B one could look at several.
Empirical data for the preferred neighbours among four groups listed as
White, Black, Hispanic and Asian in Los Angeles are given by \cite{clark}.
Corresponding Potts generalisations of the Schelling-Ising version of 
\cite{ortmanns} were published earlier \cite{schulze}. 
For Schelling models on networks, one may break the links between nodes
occupied by neighbors from different groups \cite{henry}, as done before
for other networks \cite{hohnisch}.  

In summary of this section,  cooperation of physicists with 
sociologists could have pushed research progress by many years.

\section{Opinion Dynamics}

Much of the opinion simulations, \cite{lorenz} and ch. 6 in \cite{newbook}, is  
based on the voter or majority-vote models \cite{liggett}, the 
negotiators of Deffuant et al
\cite{deffuant}, the opportunists of Hegselmann and Krause \cite{hegs}, and
the Sznajd missionaries \cite{sznajd}, the latter three originating all near
the year 2000. They check if originally randomly distributed opinions converge
towards one (``consensus''), two (``polarisation'') or more (``fragmentation'')
shared opinions. (Warning: In some fields ``polarisation'' means a non-centric
consensus, like in ferroelectrics.) See also \cite {galam}. 

The {\it voter} or majority-vote models \cite{liggett} are Ising-like: Opinions 
are $+1$ or $-1$, and at each iteration everybody follows one randomly selected 
neighbour or the majority of the neighbourhood, respectively, 
except that with a probability $q$ (which corresponds to thermal noise) it
refuses to do so.  Ref.\cite{liggett} gave a recent application. 

{\it Negotiators} of Deffuant et al \cite{deffuant} each have an opinion which
can be represented by a real number or by an integer. (Opinions on more than one
subject are possible \cite{jacob} but we first deal with one opinion only. 
Integer opinions are simplifications and often used in opinion polls when people
are asked if they agree fully, partly, or not at all with an assertion.) Each 
agent interacts during one time step with a randomly selected other negotiator. 
If their two opinions differ, each opinion shifts partly to that of the 
other negotiator, by a fraction of the difference. If that fraction is 1/2, they
agree  which is less realistic than if they nearly agree (fraction $<$ 1/2).
But if the two opinions are too far apart, they don't even start to negotiate
and their opinions remain unchanged. Thus there are no periodic boundary 
conditions applied to opinions; in contrast to real politics, the extreme Left 
and the extreme Right do not cooperate. (Axelrod studied some aspects already
earlier \cite{axel}.) 

The {\it opportunists} also talk only with people who are not too far away from
their own opinion (a real number). Each person at each time step takes the
average opinion of all the people in the system which do not differ too much
of their own past opinion. Thus in contrast to the binary interactions of
negotiators we have multi-agent interactions of opportunists. (Instead of 
opportunism one can also talk of compromise here but that word applies better
to the negotiators of Deffuant et al.)

Finally, the missionaries of the Sznajd model \cite{sznajd} try to convince
the neighbourhood of their own opinion. They succeed if and only if two 
neighbouring missionaries agree among themselves; then they force this opinion
onto their neighbourhood (i.e. on six neighbours for a square lattice). For
only two opinions on the square lattice, one has a phase transition depending on
the initial fraction of randomly distributed opinions: The opinion which 
initially had a (small) majority attracts the whole population to its side. 
In one dimension no such transition takes place \cite{sznajd}, just as 
in the Ising model. For a review of Sznajd models see \cite{sznajd2}. 

A review of both negotiators and opportunists was given by Lorenz \cite{lorenz}.
More information on missionaries, negotiators and opportunists, also for
more than one subject one has an opinion on \cite{jacob}, are given in 
\cite{newbook}. 

\bigskip

A connection between the above opinion dynamics and econophysics are the 
modifications of the usual Ising model to simulate tax evasion. Spin up 
corresponds to honest tax payers, and spin down to people who cheat on their
income tax declaration. (I am not an experimental physicist). For $T > T_c$
without any modification, half of these Ising tax payers cheat. However, if 
every year with probability $p$ the declarations are audited and fraud is
detected, and if discovered tax cheaters then become honest for $k$ consecutive 
years, the fraction of tax cheaters not surprisingly goes down towards zero 
for increasing $p$ and/or $k$ \cite{zaklan,limahohn}, also on various 
networks. Journal of Economic Psychology even plans a special issue on 
tax evasion. According to www.taxjustice.net, July 2012, hundreds of 
Giga-Dollars of taxes due are not paid world-wide each year.

\section{Wars and Lesser Evils} 
Reality is not as peaceful as a computer simulation, and World War II was
presumably the most deadly of all wars, with World War I far behind.

The ``guilt'' for World War I was hotly debated for decades; in contrast to
many books and articles, the Versailles peace treaty of 1919 did not blame 
only Germany for this war. Richardson \cite{rich} used simple diffential
equations to explain this and other wars as coming from the other side's
armaments and  own dissatisfaction with the status quo, while the cost of
armaments and war pushes for peace. Later work \cite{hermann} used human beings
to simulate ten leaders during the weeks before World War I; these simulators
made more peaceful decisions than the politicians of July 1914. Another
work simulated only the German emperor and the Russian tsar \cite{pool}
while \cite{holsti} for a more complex study used a supercomputer of that time.
Historians criticised that work because it relied on outdated history books
which explained the war more by accidents than by intentions. Except for 
Richardson \cite{rich} this early work is barely cited in physics journals.

The creation of the anti-Hitler coalition of World War II was simulated 
by Axelrod and Bennett \cite{axel}. It contains numerous parameters describing
the properties of European countries and their interrelations and thus is
difficult to reproduce. Galam \cite {galamab} has more fundamental 
criticism of that work. 

\begin{figure}[hbt]
\begin{center}
\includegraphics[angle=-90,scale=0.48]{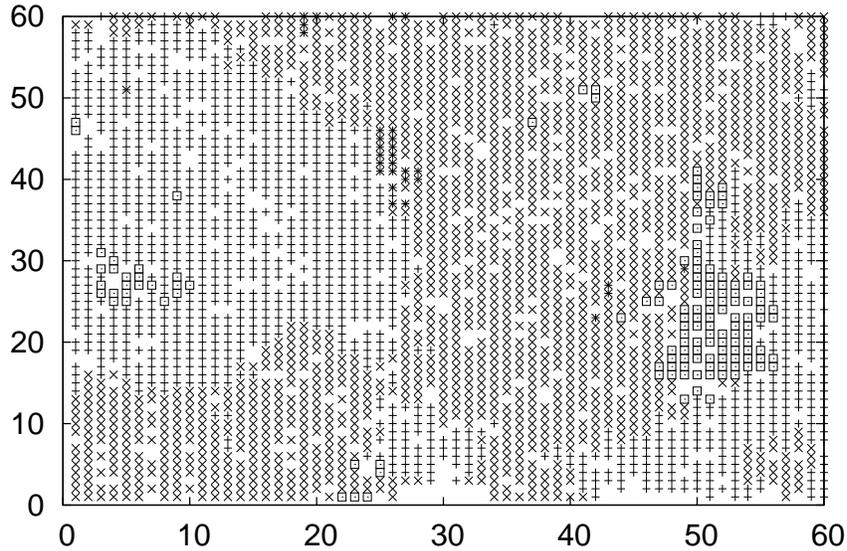}
\end{center}
\caption{Corner of a $300 \times 300$ square lattice with 1 \%  Watts-Strogatz
rewiring at intermediate times, with egoists (*), ethnocentrics (+), altruists
(x) and cosmopolitans (square) \cite{hadz2}, showing domain formation within a 
large population. The last three groups help other by an amount proportional to 
the amount of help they got from them during the five preceding time steps.}
\end{figure}

\begin{figure}[hbt]
\begin{center}
\includegraphics[angle=-90,scale=0.48]{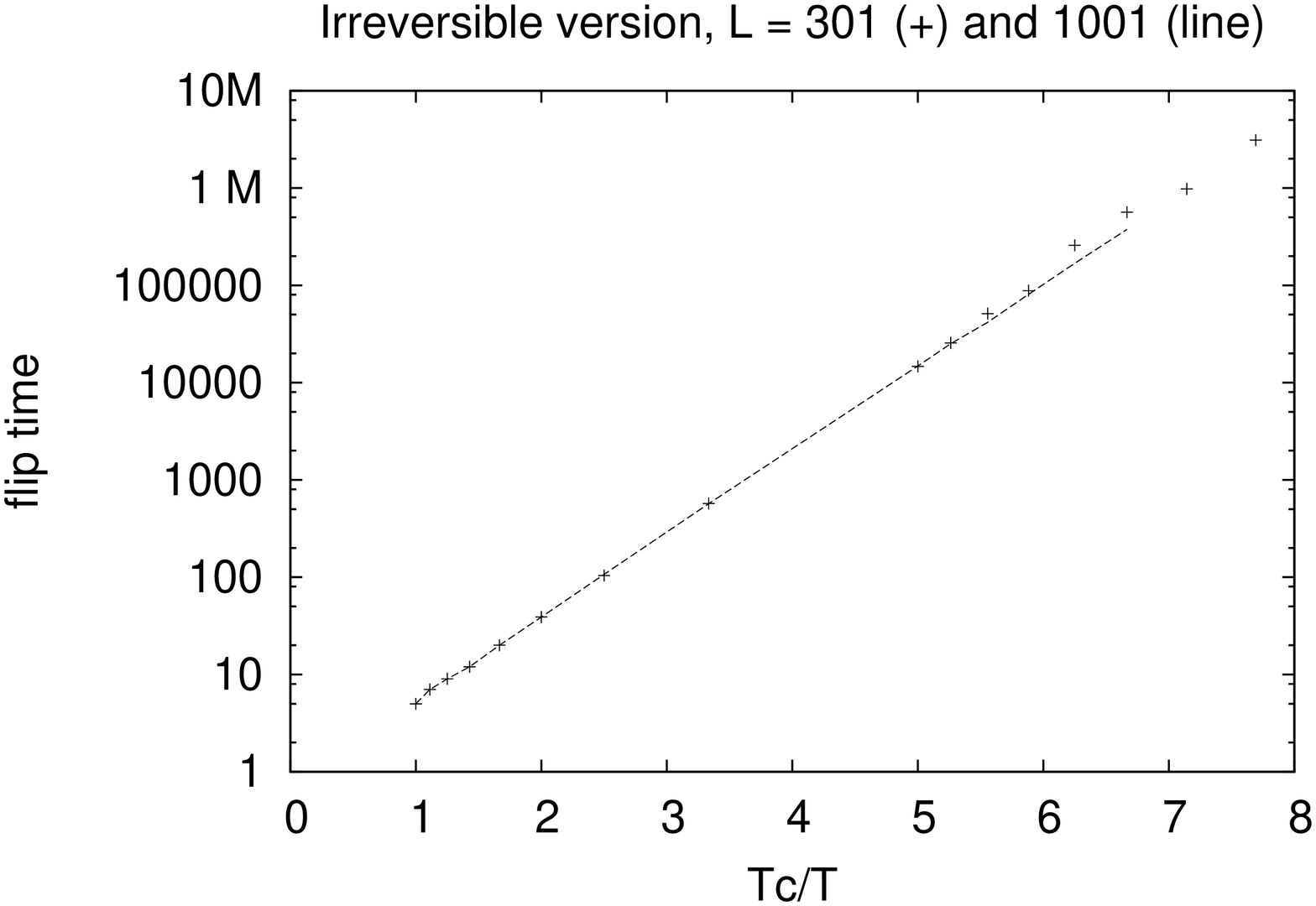}
\end{center}
\caption{Time until revolution versus $T_c/T$. A straight line here means an
Arrhenius law; the line in the figure corresponds to $L = 1001$, the plus signs
to $L = 301$; also $L = 3001$ (for higher temperatures only) gave about the
same flip times. From \cite{kindler}. 
}
\end{figure}

The decay of Yugoslavia led to the most murderous wars in Europe after World 
War II, particularly in Bosnia-Hercegovina 1992-1995 including genocide. The
emergence of local fighting between three groups was simulated by Lim et al.
\cite{lim} using a Potts model. Intergroup fighting is possible if the regions
where one group dominates are neither too small not too large. Lim et al. apply 
their model to the Bosnia-Hercegovina war but neglect the outside initiation 
and influence from Belgrade (Serbia) and Zagreb (Croatia) in that war 
\cite{courts}. Such influence was partially
taken into account in a linguistic simulation later \cite{hadz}. Fig.1
shows computer simulations of egoist, ethnocentric, altruistic and cosmopolitan
behaviour in a population \cite{hadz2}.  Often the Yugoslavia wars 
are described as ethnic. ``Ethnic'', defined e.g.  through language, 
religion \cite{ausloos} , history, biology (race, ``blood'', DNA) is now part 
of international law through UN resolutions since 1992 against 
``ethnic cleansing''. Ethnicity is often construed or even imposed on people
\cite{ethnic}, but the deaths and losses of homeland are real.

How to win a war is another question. Kress \cite{kress} reviewed modern 
simulation methods of war and other armed conflicts. Mongin \cite{mongin} 
applied game theory to a military
decision made by Napol\'eon before the battle of Waterloo, two centuries ago.
Mongin concludes that the decision was correct; nevertheless Napol\'eon lost
and ABBA won. 

Revolutions may lead to war; the ones of 2011  in Tunesia and Egypt did not 
and inspired an Ising model for revolutions \cite{kindler}. Ising spins point 
up for people wanting change, and down for staying with the government. They
are influenced by an up-field proportional to the number of up spins \cite{bornholdt},
and by a random local down field measuring the conservative tendency of each 
individual ``spin''. Initially all spins are down, and they flip irreversibly 
up by heat bath kinetics. After some time, spontaneously through thermal 
fluctuations and without the help of initial revolutionaries like Lenin and 
Trotzky, enough revolutionary opinions have developed to flip the magnetisation
from negative to positive values. This time obeys an Arrhenius law proportional 
to exp({\rm const}/T), Fig.2.

The US War of Independence is the reason that the British game of football is 
called soccer in the USA. \cite{football} confirmed that single stars in the 
team are not enough; one needs multi-player team coordination. And 
\cite{janke} found that goals are not achieved randomly in time; scoring one
goal increases the chances for the scoring team to make another goal soon 
thereafter. (See \cite{sire} for lesser games.) The sad state of football in the
author's home town, mentioned by the New York Times, prohibits further 
discussion. 

\section{Citations}

The Science Citation Index (www.newisiknowledge.com) is expensive but useful.
One can find which later journal articles cited a given paper or book, provided
the company of the Institute of Scientific Information subscribes to that
journal. Since the end of the 1960s I check my citations on it. But one should
be aware of the fact that cited books are listed not at all or only under the
name of the first author, while cited journal articles are also listed under
the name of the further authors. And citing books are ignored completely.
For example, the most cited work of the late B.B. Mandelbrot is his 1982 book:
The Fractal Geometry of Nature. The nearly 8000 citations can be found on the
Web of Science under ``Cited Reference Search'', but if after ``Search`` and
``Create Citation Record'' one gets his whole list of publications, ranked by 
the number of citations, the book is missing there and a journal article 
with much less citations heads the ranked list. 

Thus it is dangerous to determine scientific quality by the number of 
citations or by the Hirsch index (h-index with ``Create Citation Record'')
\cite{hirsch} as long as books are ignored. An author who knows  the own 
books and their first authors can include the cited books on ``Cited Reference
Search'', but automated citation counts like the h-index ignore them. If 
scholars get jobs or grants according to their h-index of other book-ignoring 
citation counts, this quality criterion will discourage them to write books,
and push them to publish in Science or Nature. This is less dangerous for
physics than for historiography, but seldomly mentioned in the literature.

(To determine the h-index, the ranked list of journal articles produced by
``Create Citation Record'' starts with the most-cited paper with $n_1$ 
citations, then comes the second-most cited one with $n_2$ citations, and
in general the $r$-ranked article with $n_r$ citations. Thus $n_r$ decreases 
with increasing rank $r$.The h-index is that 
value of $r$ for which $n_r=r$.)  

Physicists have systematically analysed citation counts (instead of merely 
counting their own and those of their main enemies) at least since Redner
1998 \cite{redner}. His work was cited more than 500 times, about half as
much as the later h-index \cite{hirsch}.  Recent papers by physicists 
deal with 
universality in citation statistics  \cite{cast}, 
the tails of the citation distribution \cite{sol}, 
co-author ranking \cite{ausloos2}, 
allocation among coauthors \cite{galamsci}, and
clustering within citation networks \cite{ren}. 

Other criticism of quality measures via citations are better known \cite{sil}. 
One should not forget, however, that experienced scientists often have to grade 
works of their students; are these evaluations more fair than citation counts? 
And what about peer review? My own referee reports are infallible, but those for
my papers are nearly always utterly unfair. Measuring quality is difficult,
but the one who evaluates others has to accept that (s)he is also evaluated 
by others. As US president and peace Nobel laureate Jimmy Carter said three 
decades ago: ``Life is unfair.''  

And as the citation lists for Redner \cite{redner} and Hirsch \cite{hirsch}
show, this problem is truely interdisciplinary. 

\section{Critique of Mean Field Theories}
This section explains mean field theory for readers from outside statistical
physics, as well as its dangers.

If you want to get answers by paper and pencil, you can use the mean field
approximation (also called molecular field approximation), which in economics
correponds to the approximation by a representative agent. Let us take the Ising
model on an $L \times L$ square lattice, with spins (magnetic moments, binary 
variables, Republicans or Democrats) $S_i = \pm 1$ and an
energy $$E = -J \sum_{<ij>} S_i S_j  - H \sum_i S_i       $$
where the first sum goes over all ordered pairs of neighbor sites $i$ and $j$.
Thus the ``bond'' between sites $A$ and $B$ appears only once in this sum, and 
not twice. The second sum proportional to the magnetic up field $H$ runs over 
all sites of the system.  We approximate in the
first sum the $S_j$ by its average value, which is just the normalised
magnetisation $m = M/L^2 = \sum_i S_i/L^2$. Then the energy is
$$E = -J \sum_{<ij>} S_i m - H \sum_i S_i = -H_{eff} \sum_i S_i$$
with the effective field

$$H_{eff} = H + J\sum_j m = H + Jqm $$
where the sum runs over the $q$ neighbours only and is proportional to
the magnetization $m$. Thus the energy $E_i$ of spin $i$ no longer is coupled
to other spins $j$ and equals $\pm H_{eff}$. The probabilities $p$ for up and
down orientations are now

$$p(S_i=+1) = \frac{1}{Z} \exp(H_{eff}/T) ; \quad p(S_i=-1) = \frac{1}{Z} 
\exp(-H_{eff}/T) $$
with 

$$Z = \exp(H_{eff}/T) + \exp(-H_{eff}/T) $$ 
and thus

$$ m = p(S_i=+1) -  p(S_i=-1) = \tanh(H_{eff}/T) = \tanh[(H + Jqm)/T]$$
with the function tanh$(x) = (e^x - e^{-x})/(e^x+ e^{-x})$. This implicit
equation can be solved graphically; for small $m$ and $H/T$, tanh$(x) = x-x^3/3 +
\dots$
gives

$$H/T = (1-T_c/T)m + \frac{1}{3} m^3 + \dots ; \quad T_c = qJ$$
related to Lev Davidovich Landau's theory of 1937 for critical phenomena
($T$ near $T_c$, $m$ and $H/T$ small) near phase transitions.

All this looks very nice except that it is wrong: In the one-dimensional
Ising model,  $T_c$ is zero
instead of the mean field approximation $T_c=qJ$. The larger the number of
neighbours and the dimensionality of the lattice is, the more accurate is
the mean field approximation. Basically, the approximation to replace $S_iS_j$
by an average $S_im$ takes into account the influence of $S_j$ on $S_i$ but not
the fact that this $S_i$ again influences $S_j$ creating a feedback.

Thus, instead of using mean field approximations, one should treat each spin 
(each human being, ...) individually and 
not as an average.  Outside of physics such simulations of many individuals are 
often called ``agent based'' \cite{billari}, presumably the first one was the
Metropolis algorithm
published in 1953 by the group of Edward Teller, who is historically known from
the US hydrogen bomb and Strategic Defense Initiative (Star Wars, SDI).

Of course, physicists are not the only ones who noticed the pitfalls of 
mean field approximations. For example, a historian \cite{siegel} years ago
criticised political psychology and social sciences: ``There is no collective 
individual'' or ``generalised individual''. And common sense tells us that
no German woman gave birth to 1.4 babies, even though this is the average
since about four decades. 
A medical application is screening for prostate cancer. Committees in USA,
Germany and France in recent months recommended against routine screening
for PSA (prostate-specific antigen) in male blood, since this simple test
is neither a sufficient nor a necessary indication for cancer. However, 
I am not average, and when PSA concentration doubles each semester while
tissue tests called biopsies fail to see cancer, then relying on PSA warnings
is better than relying on averages. 

\section{Conclusion}

Inspite of much talk since decades about the need for interdisciplinary 
research, the bibliographies on the same subject by authors from different
disciplines do not show as much overlap as they should (and as they do in 
citation analysis). Perhaps the present (literature) review helps to improve 
this situation. 

\bigskip
T. Hadzibeganovic, M. Ausloos, S. Galam, K. Kulakowski, S. Solomon helped
with this manuscript.

\end{document}